%% file: main.tex
\documentclass[sigplan,10pt]{acmart}
\renewcommand\footnotetextcopyrightpermission[1]{}

\AtBeginDocument{
  }

\usepackage{xspace}
\newcommand{\coyote}[0]{Coyote v2\xspace}

\usepackage{bbding}
\usepackage{graphicx}
\usepackage{fontawesome}

\usepackage[newfloat,frozencache,cachedir=./minted-cache]{minted}

\usepackage{caption}
\newenvironment{code}{\captionsetup{type=listing}}{}
\SetupFloatingEnvironment{listing}{name=Code}

\begin{document}

\title{\coyote: Raising the Level of Abstraction for Data Center FPGAs}

\author{Benjamin Ramhorst}
\authornote{Equal contribution.}
\email{benjamin.ramhorst@inf.ethz.ch}
\affiliation{
\institution{ETH Zurich}
\city{Zurich}
\country{Switzerland}
}

\author{Dario Korolija}
\authornotemark[1]
\authornote{Work done while at ETH Zurich.}
\email{dario.korolija@amd.com}
\affiliation{
\institution{AMD Research}
\city{Zurich}
\country{Switzerland}
}

\author{Maximilian Jakob Heer}
\email{maximilian.heer@inf.ethz.ch}
\affiliation{
\institution{ETH Zurich}
\city{Zurich}
\country{Switzerland}
}

\author{Jonas Dann}
\email{jonas.dann@inf.ethz.ch}
\affiliation{
\institution{ETH Zurich}
\city{Zurich}
\country{Switzerland}
}

\author{Luhao Liu}
\email{luhliu@student.ethz.ch}
\authornote{Also at University of Tokyo.}
\affiliation{
\institution{ETH Zurich}
\city{Zurich}
\country{Switzerland}
}

\author{Gustavo Alonso}
\email{alonso@inf.ethz.ch}
\affiliation{
\institution{ETH Zurich}
\city{Zurich}
\country{Switzerland}
}

\renewcommand{\shortauthors}{Ramhorst, Korolija et al.}

\begin{abstract}
In the trend towards hardware specialization, FPGAs play a dual role as accelerators for offloading, e.g., network virtualization, and as a vehicle for prototyping and exploring hardware designs. 
While FPGAs offer versatility and performance, integrating them in larger systems remains challenging. Thus, recent efforts have focused on raising the level of abstraction through better interfaces and high-level programming languages. Yet, there is still quite some room for improvement. In this paper, we present Coyote v2, an open source FPGA shell built with a novel, three-layer hierarchical design supporting dynamic partial reconfiguration of services and user logic, with a unified logic interface, and high-level software abstractions such as support for multithreading and multi-tenancy. Experimental results indicate Coyote v2 reduces synthesis times between 15\% and 20\% and run-time reconfiguration times by an order of magnitude, when compared to existing systems. We also demonstrate the advantages of Coyote v2 by deploying several realistic applications, including HyperLogLog cardinality estimation, AES encryption, and neural network inference. Finally, Coyote v2 places a great deal of emphasis on integration with real systems through reusable and reconfigurable services, including a fully RoCE v2-compliant networking stack, a shared virtual memory model with the host, and a DMA engine between FPGAs and GPUs. We demonstrate these features by, e.g., seamlessly deploying an FPGA-accelerated neural network from Python. 

\end{abstract}





\settopmatter{printfolios=true}
\settopmatter{printacmref=false}

\maketitle
\pagestyle{plain}

\input{sections/introduction}
\input{sections/background}

\input{sections/coyote}

\input{sections/evaluation}
\input{sections/conclusion}

\begin{acks}
We would like to thank AMD for the donation of the Heterogeneous Accelerated Compute Cluster (HACC) which was
used for the development of this project. The
work was funded in part through an unrestricted grant from
AMD.
\end{acks}

\bibliographystyle{ACM-Reference-Format}
\bibliography{main}


\end{document}

%% file: sections/introduction.tex
\section{Introduction}
Limitations on CPU performance scaling~\cite{conf/isca/EsmaeilzadehBASB11} have led to hardware specialization in data centers as a way to meet the performance demands of a diverse set of workloads~\cite{catapult, journals/cacm/AbadiAABBBBCCDD22}.
Driven mostly by  machine learning workloads, graphics processing units (GPUs), tensor processing units (TPUs), and a variety of other specialized accelerators (e.g., AWS Inferentia~\cite{amazon_inferentia}, Microsoft Maia 100~\cite{microsoft_maia}) are used to overcome the limitations of general-purpose CPUs \cite{Decline-CPU}.
In this context, Field-Programmable Gate Arrays (FPGAs) have become ubiquitous in data centers~\cite{fpgas_in_the_cloud, fpgas_in_the_cloud_2, fpgas_in_the_cloud_3}. Examples of production systems leveraging FPGAs include Microsoft AzureBoost~\cite{microsoft_boost} (formerly Catapult \cite{catapult}), Amazon AQUA~\cite{aws_aqua}, Alibaba Fidas~\cite{alibaba_fidas} and PolarDB \cite{X-Engine}, and Baidu Kunlun~\cite{baidu_kunlun}.  In these deployments, FPGAs are used either as accelerators for network functions~\cite{azure_fpga_nic, alibaba_fidas, catapult} or user applications~\cite{aws_f2, aws_aqua, baidu_kunlun}, or as prototyping platforms for next-generation accelerators and systems \cite{microsoft_brainwave, microsoft_brainwave_2}. In research, FPGAs are used to explore acceleration and offloading of, e.g., ML and database workloads~\cite{Jang_SmartInfinity, Hong_DFX, FAERY, jonas_database, farview}, networking functions~\cite{accl, Lin_SuperNIC, wang_fpganic, strom_rdma}, or cache coherency~\cite{enzian}. 

While FPGAs offer versatility and performance, integrating them in larger systems is challenging~\cite{fpgas_big_data, fos}. This happens because FPGAs are used as raw devices where every project needs to develop from scratch all the services it might need (networking, I/O, communication with the host, etc.) with very little support from either vendors or open-source systems providing such functionality. 
A recent study~\cite{harmonia} showed that approximately 75\% of the total development effort is spent on infrastructure development rather than on the application itself. Investment often lost in the next use case as the infrastructure is usually tightly integrated with the application for performance reasons and nearly impossible to reuse in other designs or other FPGAs. 

To address this issue, recent work, both commercial~\cite{catapult, sdaccel, oneapi, vitis, open_fpga, aws_f2} and academic~\cite{harmonia, miliadis, coyote, fos, optimus, amorphos, feniks} has focused on improving FPGA \emph{shells}. An FPGA shell provides abstractions and infrastructure to access shared resources (e.g., host and card memory, networking stacks, board management). Additionally, shells can include features such as memory virtualization, run-time reconfiguration, or process isolation. However, (see Section~\ref{sec:motivation}), these shells still lack key features that would allow the deployment of more complex workloads and integration with existing systems and other accelerators.

In this paper, we present \coyote, an open-source FPGA shell\footnote{GitHub repository: https://github.com/fpgasystems/Coyote} which significantly raises the level of abstraction for FPGAs and provides interfaces similar to those found in other accelerators (e.g., GPU, DPU). \coyote has been designed starting from  existing open-source resources and FPGA IPs~\cite{coyote,accl}. It is already in use and regularly maintained with contributions from academia and industry, well documented and tested, and proven to be superior in many aspects even when compared with commercial alternatives. 

\begin{table*}[t]
\caption{Comparison, with regard to the proposed requirements, of previous FPGA shells. }
\label{tab:related_work}
\scriptsize

\begin{tabular}{ccccccccc}
Shell & Services & \begin{tabular}[c]{@{}c@{}}Service\\ reconfig.\end{tabular} & \begin{tabular}[c]{@{}c@{}}Shared \\ virtual memory\end{tabular} & \begin{tabular}[c]{@{}c@{}}\textbf{Multiple}\\ reconfigurable\\ applications\end{tabular} & \begin{tabular}[c]{@{}c@{}}Multi-\\ threading\end{tabular} & Application interface & Interrupts & Open-source \\ \hline
Microsoft Catapult~\cite{catapult} & \faCheckSquare & \faSquareO & \faSquareO & \faSquareO & \faPlusSquareO & Card (single) & \faSquareO & \faSquareO \\
Xilinx SDAccel~\cite{sdaccel} & \faSquareO & N/A & \faSquareO & \faSquareO & \faSquareO & Card (single) & \faCheckSquare & \faSquareO \\
Intel OneAPI~\cite{oneapi} & \faSquareO & N/A & \faPlusSquareO & \faSquareO & \faSquareO & Host, card (single) & \faSquareO & \faSquareO \\
Vitis XRT Shell~\cite{vitis} & \faSquareO & N/A & \faSquareO & \faSquareO & \faSquareO & Host, card (single) & \faCheckSquare & \faSquareO \\
Open FPGA Stack~\cite{open_fpga} & \faSquareO & N/A & \faSquareO & \faSquareO & \faSquareO & Host, card (single) & \faCheckSquare & \faCheckSquare \\
Amazon AWS F2~\cite{aws_f2} & \faSquareO & N/A & \faSquareO & \faSquareO & \faSquareO & Host, card (single) & \faCheckSquare & \faSquareO \\ \hline
Feniks~\cite{feniks} & \faCheckSquare & \faSquareO & \faSquareO & \faCheckSquare & \faSquareO & Host, card, net (single) & \faSquareO & \faSquareO \\
AmorphOS~\cite{amorphos} & \faSquareO & N/A & \faSquareO & \faCheckSquare & \faSquareO & Card (single) & \faSquareO & \faCheckSquare \\
OPTIMUS~\cite{optimus} & \faSquareO & N/A & \faPlusSquareO & \faSquareO & \faPlusSquareO & Host (single) & \faSquareO & \faCheckSquare \\
FOS~\cite{fos} & \faPlusSquareO & \faSquareO & \faSquareO & \faCheckSquare & \faSquareO & Card (multiple) & \faSquareO & \faCheckSquare \\
Coyote~\cite{coyote} & \faCheckSquare & \faSquareO & \faCheckSquare & \faCheckSquare & \faSquareO & Host, card, net (single) & \faSquareO & \faCheckSquare \\

TaPaSCo~\cite{tapasco} & \faSquareO  & N/A  & \faSquareO & \faSquareO & \faSquareO & Host, card (single) & \faSquareO & \faCheckSquare \\

Miliadis \emph{et al.}~\cite{miliadis} & \faCheckSquare & \faSquareO & \faCheckSquare & \faCheckSquare & \faSquareO & Card (multiple) & \faSquareO & \faSquareO \\
Harmonia~\cite{harmonia} & \faCheckSquare & \faSquareO & \faSquareO & \faCheckSquare & \faSquareO & Host, card, net (single) & \faSquareO & \faSquareO \\ \hline
\textbf{\coyote} & \faCheckSquare & \faCheckSquare & \faCheckSquare & \faCheckSquare & \faCheckSquare & Host, card, net (multiple) & \faCheckSquare & \faCheckSquare 
\end{tabular}

\medskip

\faCheckSquare: Supported; \faPlusSquareO: Partially supported; \faSquareO: Not supported. \\ First group: commercial products; second group: research projects. Ordered chronologically from earliest to latest.
\end{table*}

The contributions of the paper are as follows: 
\begin{itemize}
    \item A novel, three-layer, hierarchical and modular hardware design, enabling easier portability, up to 20\% reduced synthesis times, and run-time reconfiguration of both services and applications. Service reconfiguration is an order of magnitude faster than previous approaches, which require taking the device off-line.
    \item A unified user interface facilitating the deployment of  applications, while enabling higher throughput and tighter interaction between the host CPU and the FPGA, as well as with other heterogeneous components within the system (e.g., GPUs).
    \item Several software abstractions enabling hardware sharing and reduced idle time. We show this through a multithreaded Advanced Encryption Standard (AES) block, reducing idle time up to 7x over the baseline.   
    \item Reusable and shared services to facilitate integration of the FPGA in larger systems. \coyote's services we use as examples include a fully RoCEv2-compatible RDMA stack running over a switched network and compatible with commodity hardware (e.g., Mellanox, BlueField), a modular memory management unit (MMU) with support for variable page size (e.g. 1GB huge pages) and shared virtual memory with other accelerators (e.g., direct access to GPU memory), and an on-chip network traffic sniffer.
    \item We demonstrate \coyote's versatility with several realistic workloads, including AES encryption, HyperLogLog cardinality estimation, and neural network inference. We also show how \coyote can be used to deploy complex FPGA accelerators for ML and seamlessly integrate them into Python. We do this by integrating \coyote with hls4ml~\cite{GitHub_hls4ml, Duarte_hls4ml}, a widely used, open-source ML compiler for FPGAs. By leveraging \coyote with hls4ml, users can compile and deploy neural networks on FPGAs in less than 10 lines of Python code, as is commonly done on GPUs. Additionally, due to \coyote's high-performance design, the inference is an order of magnitude faster, compared to the hls4ml baseline, with comparable resource utilization. This demonstates \coyote's ability to provide higher level abstractions with no overhead.
\end{itemize}

%% file: sections/background.tex
\section{Background \& Motivation}
\label{sec:background}

In the following, we discuss the necessary background and related work as well as motivate the design of \coyote. Table~\ref{tab:related_work} summarizes a feature comparison with related work. 

\subsection{Related Work}

\textbf{FPGA shells and OS abstractions:} Conventional FPGA shells typically provide the minimum functionality needed to communicate with the FPGA, upload bitstreams, and start and stop applications running on them. This is often enough in environments where the FPGA is used for, e.g., circuit prototyping or embedded systems. In data centers and the cloud, this is hardly sufficient and many more sophisticated shells have been developed for the purpose. Such shells implement shared and reusable services, such as memory controllers (HBM, DDR), networking stacks (RDMA, TCP/IP), or  access to host memory. They might also include memory virtualization~\cite{miliadis, coyote} or run-time application reconfiguration~\cite{harmonia, miliadis, coyote, fos}. These shells represent a significant step forward but each system provides a range of somewhat arbitrary features. Several are proprietary and not available while others are academic contributions not being maintained. For instance, Coyote~\cite{coyote}, which we use as baseline and starting point for \coyote, implements several OS abstractions such as unified virtual memory, networking, memory striping, multi-tenancy and run-time reconfiguration. However, as discussed below, its user interfaces are not generic enough to support various use cases. The service layer (e.g., networking or memory management) cannot be reconfigured without rebooting the FPGA and it lacks support for multi-threading. Similarly, Harmonia~\cite{harmonia} focuses on portability across FPGAs by using \emph{Reusable Building Blocks} (services), for host, memory and network access. But it lacks a shared virtual memory model, run-time reconfiguration of services, better abstractions and interfaces, and support for multi-threading. Miliadis \emph{et al.}~\cite{miliadis} have proposed an FPGA framework that virtualizes I/O interfaces for multiple tenants, focusing on isolation, scalability and reconfiguration. It lacks direct data streams to the host, support for networking, and does not allow run-time reconfiguration of services. Neither ~\cite{harmonia} nor ~\cite{miliadis} are open-source, limiting the ability of users to tailor or build systems on top of them. FOS~\cite{fos} facilitates the deployment of multiple, reconfigurable user applications through a standardized application interface and software run-time, but lacks support for networking and memory virtualization. OPTIMUS~\cite{optimus} spatially partitions the FPGA into multiple regions to be used by different applications but these regions are not dynamically reconfigurable. AmorphOS~\cite{amorphos} places multiple applications on a single FPGA, providing virtualized access to FPGA memory but no direct access to host memory. Neither OPTIMUS~\cite{optimus} nor AmorphOS~\cite{amorphos} support complex services, such as networking stacks. TaPaSCo~\cite{tapasco} is an open‑source framework providing an automated toolflow and software run-time for constructing complete designs with access to peripherals for arbitrary applications and executing them on a variety of FPGAs. While TaPaSCo focuses on automated hardware compostion and portability by supporting many embedded and data center FPGAs, it lacks networking services, run-time reconfiguration of applications and a shared virtual memory model.

\textbf{FPGA virtualization:} A significant body of work~\cite{vfpio, fsrf, nimblock, synergy, ruan_virtualization, vital, hetero_vital} focuses on virtualizing various aspects of the FPGA, such as memory or network peripherals. These efforts typically build on top of existing shells and provide orthogonal functionality. vFPIO~\cite{vfpio} extends Coyote~\cite{coyote} to virtualize FPGA I/O ports with preemptive scheduling, making the user logic platform-independent. FSRF~\cite{fsrf} virtualizes FPGA I/O, enabling files to be directly mapped from host to FPGA virtual memory, while optimizing the MMU for each user application. Nimblock~\cite{nimblock} partitions the FPGA fabric into multiple slots with an overlay architecture. Leveraging pipeline parallelism, user applications are split into tasks which are dynamically and pre-preemptively scheduled onto the FPGA. ViTAL~\cite{vital} virtualizes a cluster of FPGAs into a single FPGA and compiles multiple applications into abstractions that are dynamically distributed onto the cluster at run-time. HETERO-ViTAL~\cite{hetero_vital} extends the work to heterogeneous FPGA clusters. Ruan \emph{et al.}~\cite{ruan_virtualization} propose an approach that enables partial reconfiguration of vFPGAs in isolated virtual machines (VMs) through a dedicated driver. 
In comparison, \coyote has direct support for multithreading and multitenancy through the notion of \emph{virtual FPGAs} and provides isolation as well as fair-sharing between them at the level of the services provided (e.g., memory, networking), thereby enabling  different applications to concurrently run at the same time on the same FPGA. 

\textbf{Commercial products:} There are as well commercial shells and products from FPGA~\cite{sdaccel, oneapi, vitis, open_fpga} and cloud~\cite{catapult, aws_f2} vendors. The underlying shells of these frameworks are usually hard to separate from the application development process and possibly the programming language that they are integrated with. Moreover, these frameworks are generally designed for specific purposes and offer limited configurability, making them ill-suited for more generic or flexible use cases. All of these systems enable access to card memory and, except Microsoft Catapult~\cite{catapult} and SDAccel ~\cite{sdaccel}, host memory. However, they lack a lot of crucial features, such as shared virtual memory, networking stacks, and run-time reconfiguration of the application and service layers. 

\subsection{Requirements \& Motivation}
\label{sec:motivation}

Although there has been significant work on raising the level of abstraction for FPGAs, practically deploying applications on FPGAs and integrating them into larger systems remains a cumbersome and slow process. To further elaborate on this claim, we present three key requirements an FPGA shell should meet to facilitate deployment in a cloud environment.

\textbf{Requirement 1 - Reusable and reconfigurable services:} Akin to a conventional OS, an FPGA shell should include the infrastructure for commonly used services, such as memory controllers (HBM/DDR), networking stacks (RDMA, TCP/IP), compression and and encryption cores, memory virtualization, etc. While many shells~\cite{harmonia, coyote, feniks, catapult} provide such services, they are static; that is they are loaded once during initialization and are not reconfigurable. 
However, realistic workloads are dynamic in nature and reconfiguring the services (e.g., switching from TCP/IP to RDMA, varying the MMU configuration, or changing the compression algorithm) should not require to reboot the FPGA, thereby interrupting service. 
To address such limitation, \coyote's three-layer hardware design natively supports run-time reconfiguration of services, enabling efficient adaptation to dynamic workloads.

An additional goal for the design was to establish the FPGA as an equal-class citizen in the landscape of data center hardware, rather than a stand-alone compute node. Therefore, the emphasis when implementing services was on compatibility and seamless interaction with other parts of the system. For this purpose, \coyote includes a highly configurable memory management unit (MMU), which seamlessly moves data between the host CPU and the FPGA. Proof of \coyote's flexible and extensible MMU is an external contribution~\cite{p2p_github} to the open-source codebase, which extended the MMU to include GPU memory and supports direct data movement between the FPGA and a GPU as proposed in~\cite{wang_fpganic}. Another example of an independent service available in \coyote is an open-source RoCE v2-compatible RDMA stack~\cite{systems_rdma}, which, when integrated with \coyote's memory model, enables out-of-the-box interaction between the FPGA and commodity network interface cards (NICs), such as Mellanox and BlueField devices. How to implement services will be later demonstrated by describing a service for capturing network traffic in real time, similar to \texttt{ibdump} or \texttt{tcpdump} on Linux.

\begin{figure}[t]
    \centering
    \includegraphics[width=.75\linewidth]{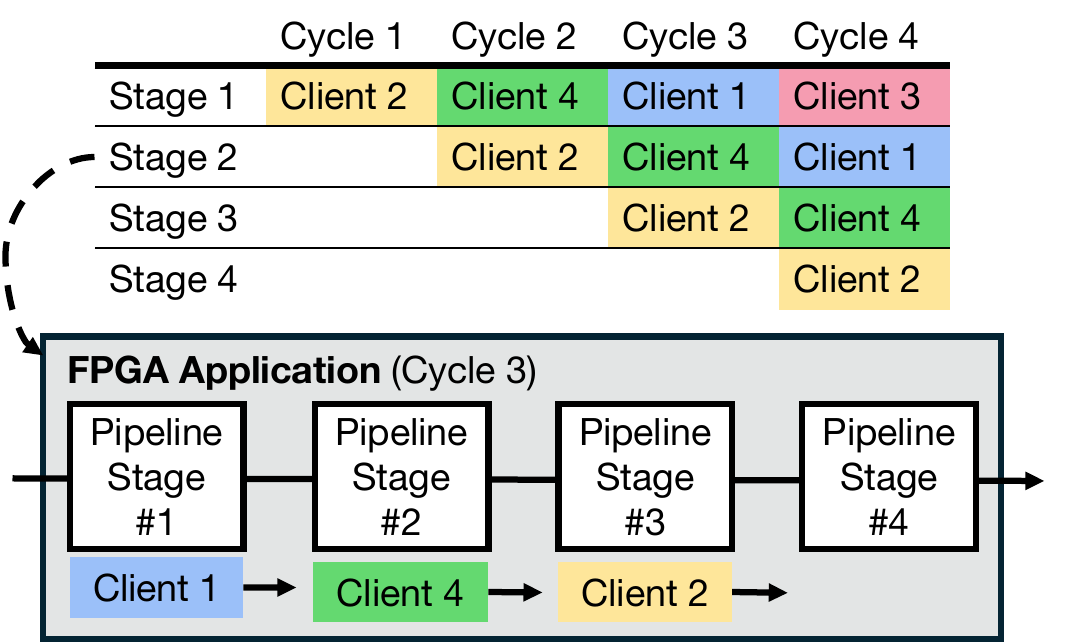}
    \caption{Example of a multi-threaded application with four pipeline stages in hardware and each stage independently processing the data of a client.}
    \label{fig:mt_example}
\end{figure}

\textbf{Requirement 2 - Multi-tenancy, partial reconfiguration and multi-threading:} FPGA hardware has been steadily improving over the past few years, often providing compute, memory and network bandwidth exceeding the requirements of a single application. For example, the recent AMD Alveo V80~\cite{v80} is equipped with over 2.5 million LUTs, 800G networking hardware and 32 GB of 810 GBps-capable HBM. Given this trend, several shells~\cite{harmonia, coyote, miliadis, fos} have proposed spatial sharing with partial reconfiguration of the user region. Following this trend, \coyote also includes the ability to deploy multiple, reconfigurable user applications, while ensuring fair access to the shared services. 

In contrast to existing work, however, \coyote provides user applications with multithreading. Certain workloads can be deeply pipelined, but still fail to achieve high throughput due to data dependencies. Examples include (i) AES Chipher Block Chaining (CBC) encryption, which encrypts text sequentially in chunks, but each chunk depends on the previously encrypted one, and (ii) LLMs, where each token depends on the previously generated token. In these cases, avoiding idle time in hardware can be achieved by servicing multiple requests simultaneously (Figure~\ref{fig:mt_example}). In our case, multithreading is a direct consequence of the improved user interfaces and, to the best of our knowledge, not available in any other shell.

\begin{figure}[t]
    \centering
    \includegraphics[width=\linewidth]{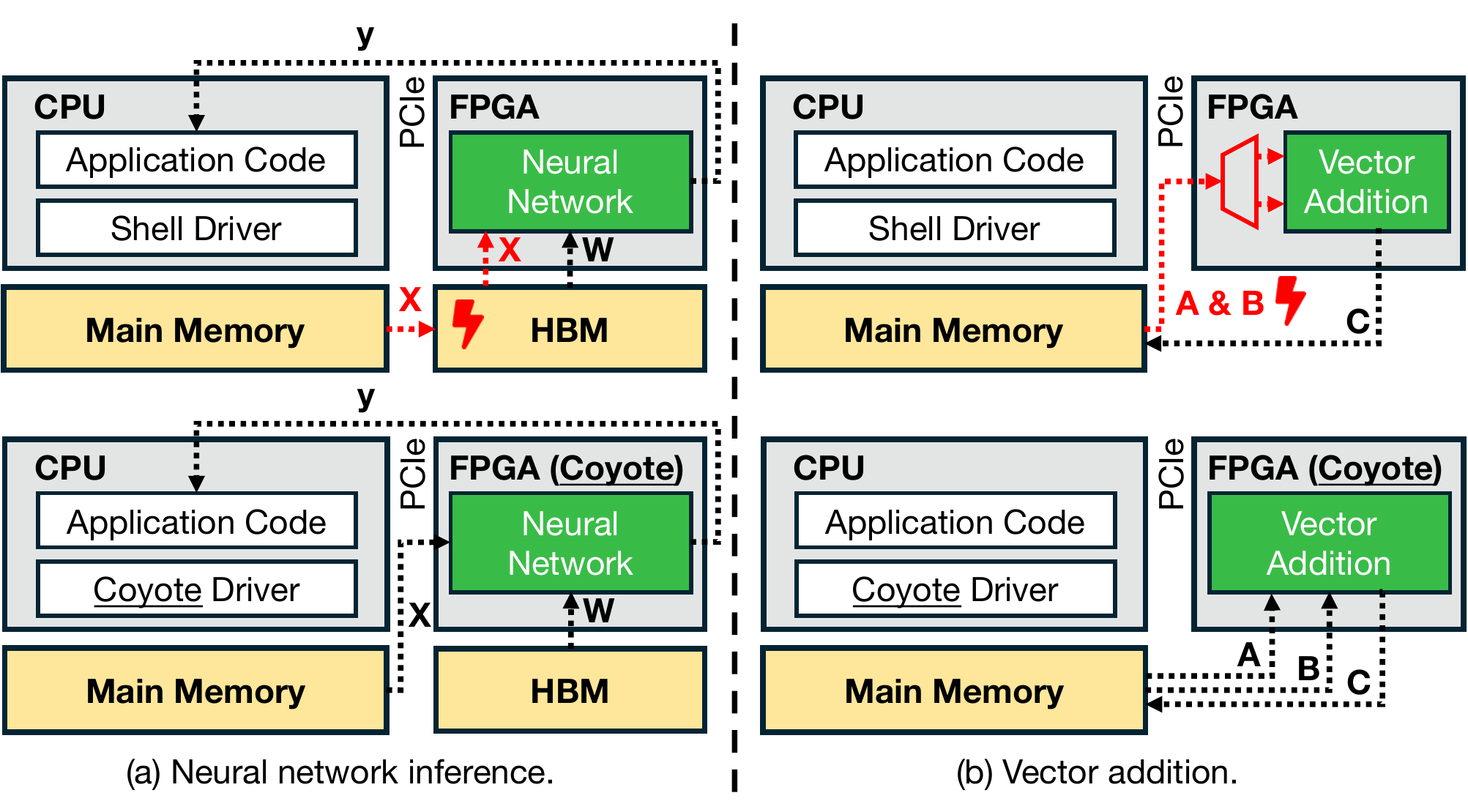}

    \caption{Limitations of existing shell interfaces and the proposed solutions with \coyote's interfaces.}
    \label{fig:interfaces_example}
\end{figure}

\textbf{Requirement 3 - Unified and generic application interfaces:} To facilitate data movement and kernel control, shells typically provide standardized interfaces. While several projects, including Coyote~\cite{coyote}, Harmonia~\cite{harmonia} and AmorphOS~\cite{amorphos} have proposed such interfaces, they are often not sufficiently generic as can be shown with two examples (Figure~\ref{fig:interfaces_example}):

\begin{itemize}
    \item Neural network inference: Similar to GPUs, FPGAs have been successfully used for neural network inference~\cite{fpgas_nn_survey, wenqi_microrec, finn}. Adopting an approach commonly found on GPUs, weights can be pre-loaded to FPGA HBM, while input data can be directly streamed from the host as requests come in. In this case, AmorphOS~\cite{amorphos} requires the input data to be first copied from host memory to FPGA HBM, before it can be processed by the application. When compared to Coyote~\cite{coyote}, which streams data directly from host memory to the user application, completely bypassing FPGA HBM, the approach taken by AmorphOS incurs a non-negligible latency penalty. Similarly, SDAccel~\cite{sdaccel} enables data streams from the host only on specific QDMA-enabled FPGA platforms. 
    \item Vector addition: For vector addition, an FPGA application should consume two (or more) vectors and produce a single result vector. Typically, vectors are streamed from host memory to the FPGA; however, most previous shells~\cite{coyote, harmonia, feniks, catapult} only provide one data stream from/to host memory, requiring the user to manually pack multiple vectors into a single stream in software and then unpack them in hardware. Doing so for applications with many inputs and outputs is cumbersome and error-prone. 
\end{itemize}

\begin{figure*}[t]
    \centering
    \includegraphics[width=\textwidth]{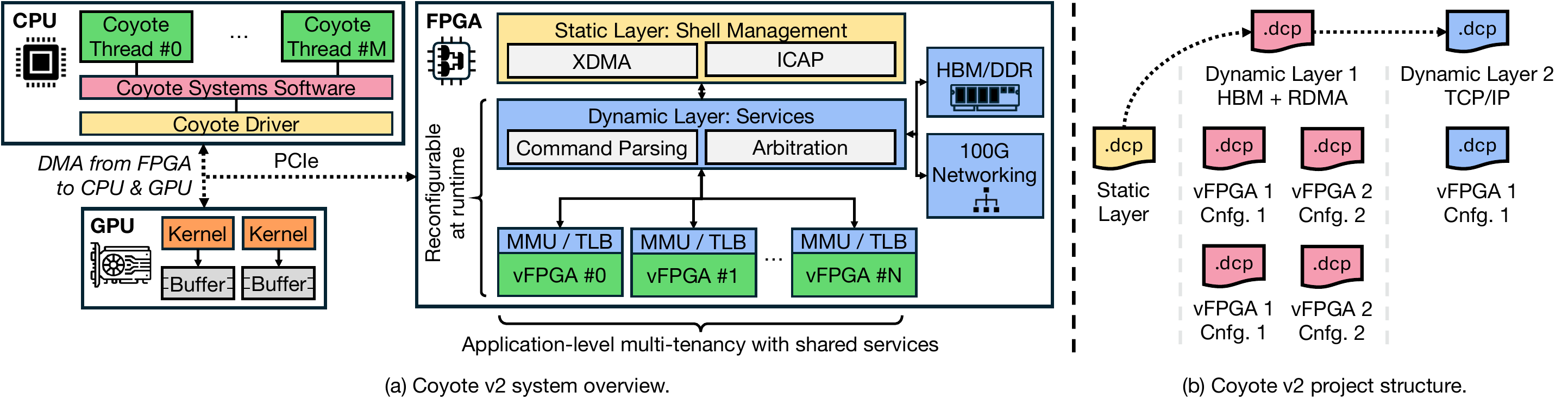}
    \caption{\coyote system overview and project structure.}
    \label{fig:system_overview}
\end{figure*}

In \coyote we define a single and generic user interface supporting multiple data streams from and to host memory, FPGA memory, and the network. Additionally, the interface allows  user applications to issue DMA requests from hardware, rather than solely relying on the host software to do so. 

Finally, only some commercial shells~\cite{sdaccel, vitis, aws_f2} provide interfaces for the user application to issue generic interrupts. However, a sufficiently generic interrupt interface is a necessity for realistic workloads, as applications can encounter various unwanted states, such as malformed data or time-outs. An interface like this would enable a closer interaction between the host CPU and the FPGA card, while also enabling the host software to take control in the case of an unexpected situation. In \coyote we include this functionality as part of the generic interface, thereby simplifying the design space without losing flexibility.

%% file: sections/coyote.tex
\section{System Overview}

In the following, we introduce \coyote (Figure~\ref{fig:system_overview}(a)), a shell addressing requirements 1-3.
\coyote is implemented using a three-layer, hierarchical and modular hardware design: (i) the static layer, (ii) the dynamic (services) layer and (iii) the application (user) layer. 
Additionally, \coyote includes (i) a user-facing, high-level software API and (ii) a device driver. We describe these layers in more detail in the following section. However, before doing so, it is important to highlight the benefits of the proposed design. 

The dynamic and the application layer together make up the so-called \emph{shell}. In contrast to previous works~\cite{coyote, feniks, fos}, a key advantage of our approach is that services are no longer part of the static layer. Instead, services are now part of the shell which can be dynamically reconfigured. By separating the services from the static layer, \coyote significantly simplifies the static layer, while also enabling independent run-time reconfiguration of both services and user applications within them. Without having to implement networking, memory controllers etc., the primary purpose of the static layer is now  only to provide a link between the host CPU and the FPGA, which can be used for data movement, control and reconfiguration. Importantly, the static layer does not process the incoming data or control signals; instead it passes them onto the upper layers, routing each request to the correct user application or service. The static layer is often card- and interconnect-dependent, since it relies on a DMA IP (e.g. XDMA, QDMA) for CPU-FPGA communication. However, by simplifying the static layer, it becomes easier to port it to other FPGAs; a fact highlighted by a recent shell, Harmonia~\cite{harmonia}. \coyote runs on a variety of AMD FPGAs (U250, U55C, U280) taking advantage of this separation of layers, also hiding the details of the hardware from the application, making such designs also portable across FPGAs. 

\section{Shell Reconfiguration}
\begin{figure}[b]
    \centering
    \includegraphics[width=.9\linewidth]{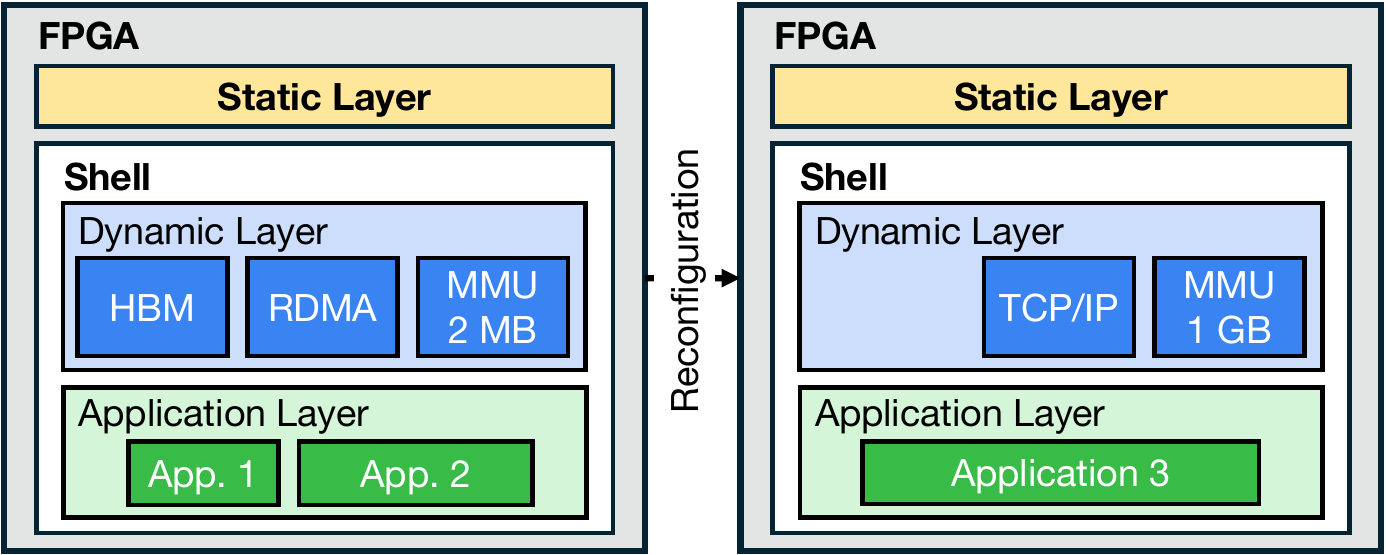}
    \caption{Shell reconfiguration example.}
    \label{fig:shell_configs}
\end{figure}

To enable shell reconfiguration, \coyote provides a floor-plan and interfaces which connect the static layer to the shell. Both the floor-plan and the interfaces are hidden from \coyote users; instead, the users simply choose the various shell configurations they would like to synthesize through compile-time parameters. A shell is fully parametrized by its services and the user applications. \coyote will then synthesize all the necessary partial bitstreams (Figure~\ref{fig:system_overview}(b)) which can dynamically be loaded onto the FPGA, without having to take the FPGA off-line. A further advantage of moving the services from the static layer to the shell layer is faster synthesis. Since the primary role of the static layer is to provide a link between the shell and the host CPU, it is not configurable. Therefore, \coyote provides a routed and locked checkpoint of the static layer for each supported FPGA, which can be linked with the shell. An example of a shell reconfiguration process, alongside common shell configurations, is shown in Figure~\ref{fig:shell_configs}.

Compared to shell reconfiguration, \coyote also includes the option to independently reconfigure individual user applications, without affecting other applications or services. This is similar to approaches proposed by prior work~\cite{coyote, harmonia, fos, miliadis} which can trigger reconfiguration of specific applications as user requests arrive, based on some scheduling policy. It is important to note that, in \coyote, user applications \emph{depend} on the shell services; that is, a full shell reconfiguration will reconfigure both the services and the user applications, while an application reconfiguration will only reconfigure the user application. This is primarily a fail-safe mechanism which prevents a running application from losing access to a service it requires. Instead, when synthesizing the hardware, an application is always linked to a shell configuration, which verifies that the services required by the application are indeed provided by the shell configuration. This approach introduces an additional layer of isolation and supports multiple privilege levels. Another clear advantage of this modular design is the ability to link new user applications against previously synthesized shell configurations, reducing synthesis times. 

In conclusion, the proposed three-layer hardware design offers greater design flexibility and reconfiguration opportunities, while still maintaining the ability to only reconfigure user applications, as proposed by previous shells~\cite{sdaccel, feniks, fos, coyote, amorphos, harmonia}. Furthermore, it simplifies the static layer, which can in turn, simplify the porting of the shell to other FPGAs~\cite{harmonia}.

\section{Static Layer}
The static layer interfaces with platform-specific hardware, providing a link to the host CPU and the logic necessary to reconfigure the shell. Additionally, the device driver is a core component of the static layer. 

\subsection{CPU-FPGA Link}
Heterogeneous CPU-FPGA systems excel at adapting to diverse workloads, by dispatching different tasks to the hardware better suited for the task. Additionally, while FPGAs offer flexibility and high compute performance, they are less suited for management tasks, which are generally better handled on the host CPU. Therefore, the static layer must provide an efficient and generic interface to handle data and control flow between the FPGA and the host CPU. \coyote uses the AMD XDMA core~\cite{xdma}, which functions as a DMA wrapper on top a hardened PCIe block on the FPGA, and importantly, can be controlled from both the FPGA and the CPU. Consequently, this allows the FPGA to initiate data transfers with no host involvement. On the other side, the static layer interacts with the shell through these  channels:

\textbf{Shell control:} Memory mapping various shell control registers (e.g., TLB control, network configuration, interrupt registers etc.)  into host memory via PCIe Base Address Registers (BARs).

\textbf{Host streaming channel:} An AXI4 stream for direct data movement between host memory and user applications. This channel reduces data transfer latency, compared to shells~\cite{amorphos, catapult, miliadis} which require a copy from host memory to card memory, before being available to the user application.

\textbf{Migration channel:} Enables migrating buffers from host memory to FPGA-side memory (HBM/DDR). Such operations are required for large buffers where transfer latency is not a concern, e.g. transferring the weights before model serving for inference. Note, this channel is only available when the shell is built with memory support; otherwise it is tied off in the dynamic layer.

\textbf{Utility channel:} This two-sided channel is primarily reserved for management operations. On one side, partial bitstreams for reconfigurations are loaded from the host via this channel, which we describe further in the following sub-section. On the other side, this channel is used for \coyote's \emph{writeback} functionality. The writeback mechanism enables efficient completion tracking by updating host memory counters when data transfers finish. This reduces unnecessary PCIe polling, thus freeing up bandwidth. While the XDMA core natively supports writeback with host-mapped counters, we extend it to all additional data services: FPGA memory and the network; all of which operate independently of the PCIe. Such functionality is crucial for high-performance systems, but is often found lacking. Finally, this channel is used to raise interrupts to the host, using the standardized MSI-X (Message Signaled Interrupts eXtended) technology, which is processed by the device driver. In a complex system like \coyote there are many sources of interrupts, such as page faults, reconfiguration completions, TLB invalidations and user-issued interrupts, a key feature for realistic workloads, commonly found missing in many shells.

\subsection{Device Driver}
\coyote's device driver is a Linux kernel component bridging user applications in software and in hardware. It manages the FPGA and its peripherals, handling memory mappings, dynamic allocations, page faults, and partial reconfiguration. The driver also initializes all user application in hardware, enabling communication from software via standard system calls like \texttt{open}, \texttt{close}, \texttt{mmap}, and \texttt{ioctl}.

\subsection{Reconfiguration Controller}
Partial reconfiguration in \coyote is managed through the Internal Configuration Access Port (ICAP)~\cite{icap}, a centralized block enabling dynamic partial reconfiguration while the rest of the FPGA remains operational. Achieving high reconfiguration speed is crucial for a realistic data center setting. Standard methods, such as AXI HWICAP~\cite{hwicap} and MCAP~\cite{amd_pr}, suffer from low throughput due to their reliance on single-word writes. To maximize performance, we implement an optimized controller that fully utilizes the ICAP bandwidth ($\sim$800MBps on AMD UltraScale+ devices). This bandwidth acheived by loading the bitstream from host memory via PCIe and a dedicated XDMA channel. Table~\ref{tab:reconfig_throughput} summarizes the differences in performance between existing reconfiguration controllers and \coyote.

\begin{table}[bt]
\caption{Reconfiguration throughput comparison.}
\begin{tabular}{ccc}
Application & \begin{tabular}[c]{@{}c@{}}Maximum \\ throughput {[}MB/s{]}\end{tabular} & Interface \\ \hline
AXI HWICAP~\cite{hwicap} & 19 & AXI Lite  \\
PCAP~\cite{amd_pr} & 128 & AXI \\
MCAP~\cite{amd_pr} & 145 & AXI \\ \hline
\coyote ICAP & 800 & AXI Stream
\end{tabular}
\label{tab:reconfig_throughput}
\end{table}

\section{Dynamic Layer}

\subsection{Memory Management}
We build upon Coyote's~\cite{coyote} shared virtual memory model, enhancing it to support arbitrary page sizes, TLB sizes and associativities.
The memory model is similar to the one commonly found in GPUs, issuing a page fault when the requested data is not in the correct memory (CPU DDR, FPGA HBM) and triggering a migration. \coyote's MMU is implemented in a hybrid manner: TLBs are implemented in on-chip SRAM, enabling fast look-ups, while the rest of the MMU is implemented in the host-side driver; that is, when a TLB miss is detected; the system falls back to the driver to obtain the physical address. A stand-out feature of \coyote is that the TLB configuration is parametrizable, allowing \coyote to be deployed on a wide range of systems. Given the ever-increasing data requirements of modern applications, particularly important is the ability to set the page size to very large huge pages (1GB), minimizing page faults. 

\coyote also abstracts the creation of any memory controllers (HBM/DDR) on the FPGA and is highly configurable, allowing developers to set options such as number of memory channels, memory clock frequency etc. Application requests to FPGA memory are also done with virtual address, with the translation being handled by the MMU. To maximize throughput, \coyote implements memory striping, partitioning data buffers across multiple HBM banks.

\subsection{Networking}
One of the key services in \coyote is BALBOA~\cite{systems_rdma}, a 100G, fully RoCE v2-compliant networking stack, that enables the deployment of a \coyote-powered FPGA in a heterogeneous networking environment. BALBOA exposes standard AXI-streaming interfaces for both data- and control-flow to the host and network, making it portable and easy to integrate in \coyote. Both data- and control-flow are routed through the user applications (vFPGAs) enabling on-datapath custom off-loads and data processing tasks, similar to SmartNICs or Data Processing Units (DPUs). Additionally, BALBOA aligns well with other abstractions from \coyote. The network stack, since it implements RDMA, operates on virtual memory addresses that are translated using \coyote's internal MMU and TLB, before writing the data to host memory through the static layer.

\subsection{Multi-tenant Fair Sharing}
To achieve fairness between multiple tenants on bandwidth-constrained links (PCIe, network), \coyote implements packetization, interleaving and a dedicated credit-based system for all data requests (from/to each vFPGA). Packetization divides transfers into manageable 4 KB chunks (default, but configurable), which enables precise control over outstanding transactions while ensuring efficient saturation of both local and remote links. The shell seamlessly splits requests of arbitrary sizes  into packets, requiring no user application involvement. Interleaving distributes limited bandwidth links using round-robin arbitration, guaranteeing equal resource allocation while preserving in-order packet handling. However, interleaving is unnecessary for FPGA HBM requests, as the significantly higher local bandwidth allows each vFPGA to utilize dedicated interfaces efficiently. 

\section{Application Layer}
\label{sec:user_layer}

The application layer consists of multiple parallel vFPGAs, which can host arbitrary user applications. We begin by describing the generic application interface, which abstracts and virtualizes data movement, control flow and network interaction for the vFPGAs. Secondly, we describe the crediting mechanism, which prevents any vFPGA from exerting back-pressure on the entire system. Finally, we present the software API, which facilitates seamless interaction with the applications described in hardware.

\subsection{Generic Application Interface}
\begin{figure}[bt]
    \centering
    \includegraphics[width=0.65\linewidth]{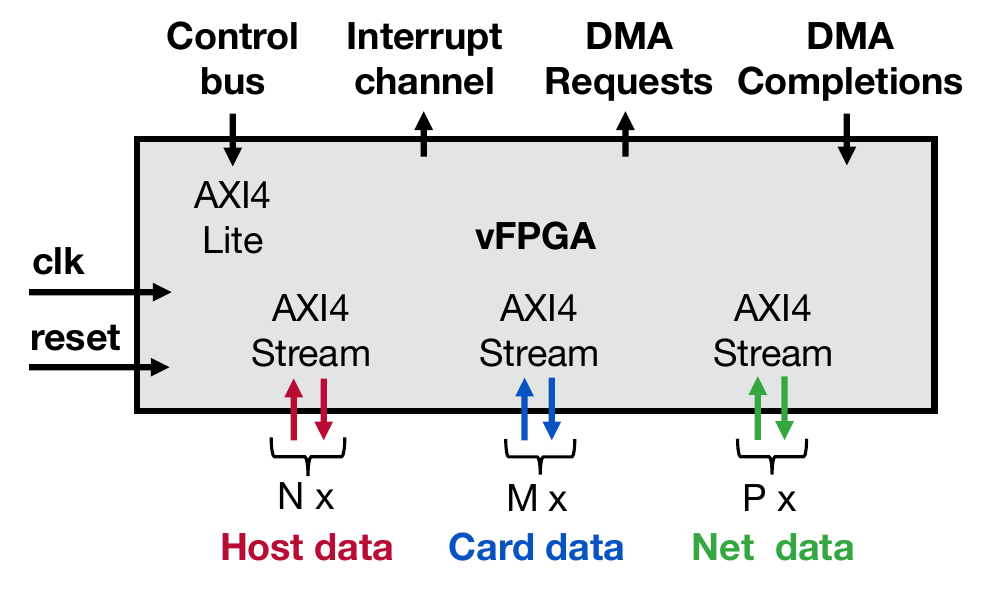}
    \caption{In- and outgoing vFPGA interfaces.}
    \label{fig:vfpga_interfaces}
\end{figure}

A generic application interfaces solves two problems: (i) the inherent complexity of accessing FPGA peripherals (host, network, card) to fetch data and (ii) portability to other platforms. By providing a standardized execution environment, developers can focus on application development and performance optimization, without having to worry about infrastructure or portability. We base our execution environment on Coyote~\cite{coyote}, but extend it to include multiple data streams and user interrupts. The interfaces, illustrated in Figure~\ref{fig:vfpga_interfaces}, are built around the industry-standard AXI specification and include:

\textbf{Control bus:} enables software control over the deployed user applications. This interface is built around an AXI4 Lite bus, which is memory-mapped for each vFPGA directly into the user space, bypassing the kernel space, to achieve lower latency. On the hardware, this interface connects to a set of control and status registers, whose functionality is application-specific and user-defined.

\textbf{Interrupt channel:} enables hardware applications to issue interrupts, with arbitrary values, to the user space. On the host, interrupts are polled using the standard Linux \texttt{eventfd} mechanism, which can trigger an interrupt callback function in the user-space.

\textbf{Parallel host interface:} multiple AXI4 streams, for data from and to host memory. Parallel streams can be used to accomodate multiple software threads with the same hardware. This concept becomes key for multi-threaded applications, as discussed in Section~\ref{sec:mt_results}.

\textbf{Parallel card interface:} multiple AXI4 streams, for data from and to card memory (HBM/DDR). Parallel streams can be used to acheive higher throughput, especially with HBM memory, as discussed in Section~\ref{sec:hbm_results}.

\textbf{Parallel network interface:} multiple AXI4 streams, for data from and to the enabled networking stacks.

\textbf{Read and write send queues:} interfaces which enable the vFPGA to trigger local and remote data transfers, without relying on host software, by specifying information such as buffer virtual address, length, type of operation (local or remote), target stream etc. Parallel interfaces for reads and writes to achieve higher performance. This interface is particularly important for accelerated applications that rely on some sort of pointer chasing. In a host-centric system, the CPU would have to manage each data transfer between memory and the FPGA. At every step, the CPU must either poll or handle interrupts to initiate subsequent data movements, leading to increased latency and wasted CPU cycles. 

\textbf{Read and write completion queues:} interfaces containing information for completed data transfers.

Key advantages of the proposed interfaces are: (i) easier deployment of applications with multiple inputs/outputs, (ii) higher HBM throughput through parallel accesses (Section~\ref{sec:hbm_results}), (iii) multi-threaded user applications (Section~\ref{sec:mt_results}), and (iv) better interaction with the host through the interrupt channel.

\subsection{Untrusted Applications and Crediting}
Since \coyote's vFPGAs house arbitrary user logic, they are assumed to be untrusted, and like OS processes, safe-guards preventing system-wide disruptions must be implemented. To do so, \coyote first implements a per-vFPGA MMU ensuring memory isolation between multiple vFPGA. Second, \coyote implements a credit system to tackle backpressure and deadlocks. As an example, consider a scenario in which a single vFPGA requests data but, upon receiving it, fails to consume it. Such scenarios have the potential to create backpressure on the entire system, disrupting the operation of other applications and the system as a whole. Similar scenarios can occur for write requests when no data to write is provided. For each vFPGA, \coyote implements a per-stream crediting mechanism, built on top of destination queues, which verifies the available credits for the specific vFPGA and data stream. Requests are only propagated to the dynamic layer when sufficient space in the queue is available. Otherwise, the request is stalled, exerting back-pressure onto the vFPGA rather than the rest of the system. Credits are replenished when previous requests are marked as complete. Crediting applies to all data requests: host, card memory and, network, with independent crediters implemented for each of the three, maximizing performance and parallelism.

\subsection{Software API}
Finally, we introduce \coyote's software API, implemented in C++ for  cross-platform compatibility, high performance, and support for low-level operations. While assigning each vFPGA to an individual host process may seem intuitive, congestion and routing constraints practically limit the number of active vFPGAs to between eight and ten. An alternative for achieving higher performance on FPGAs is leveraging pipeline parallelism, which enables independent threads to advance through the pipeline stages of the hardware application without having to wait for other threads to complete (Figure~\ref{fig:mt_example}).

To do so, we introduce \coyote threads, \emph{cThreads}, corresponding to software threads that execute in parallel on the same vFPGA pipeline, while preserving thread differentiation. Data isolation can be achieved by ensuring each thread uses a unique subset of the previously described parallel data interfaces in the vFPGA. This concept mirrors hyperthreading in modern CPUs, where multiple threads share a core to mitigate data access latency.
Each \emph{cThread} is associated with a specific vFPGA and can be used to allocate card memory, set and read control registers, trigger data movement, initiate Queue Pair (QP) numbers for RDMA connections and invoke hardware kernels. We demonstrate the simplicity of interaction with \coyote in the following code example, show-casing how an encyption application can be invoked from software.

\begin{code}
\captionof{listing}{Example \coyote code for memory allocation, control register setting and kernel launch.}
\begin{minted}{c++} 
// Create a cThread and assign it to vFPGA 0
cThread<std::any> cthread(0, getpid());

// Allocate 4KB source & destination memory 
// using huge pages (HPF)
// Also, getMem adds src and dst to the TLB
char *src = cthread.getMem({Alloc::HPF, 4096});
char *dst = cthread.getMem({Alloc::HPF, 4096});

// Some host-side processing on src and dst

// Set hardware register for encryption key
const uint64_t KEY = 0x6167717a7a767668;
cthread.setCSR(KEY, 0);

// Create SG entry for DMA transaction
sgEntry sg;
sg.local = { 
    .src_addr = src, .src_len = 4096, 
    .dst_addr = dst, .dst_len = 4096
};

// Launch the kernel 
// Specifying the source and destinaton buffer
 cthread.invoke(Oper::LOCAL_TRANSFER, &sg);
\end{minted}
\end{code}

Additionally, the software API includes the necessary functions to handle reconfiguration. To do so, it is simply required to pass a path to the partial bitstream file:

\begin{code}
\captionof{listing}{Example \coyote code for dynamic reconfiguration.}
\begin{minted}{c++} 
// Create a reconfiguration instance
cRcnfg rcnfg(0);

// Shell (dynamic + app) reconfiguration
rcnfg.reconfigureShell("/path/to/shell.bin");

// vFPGA #2 (app) reconfiguration
rcnfg.reconfigureApp("/path/to/app.bin", 2);
\end{minted}
\end{code}

\section{Case Study: Traffic Sniffer}
\label{sec:traffic_sniffer}
\begin{figure}[bt]
    \centering
    \includegraphics[width=\linewidth]{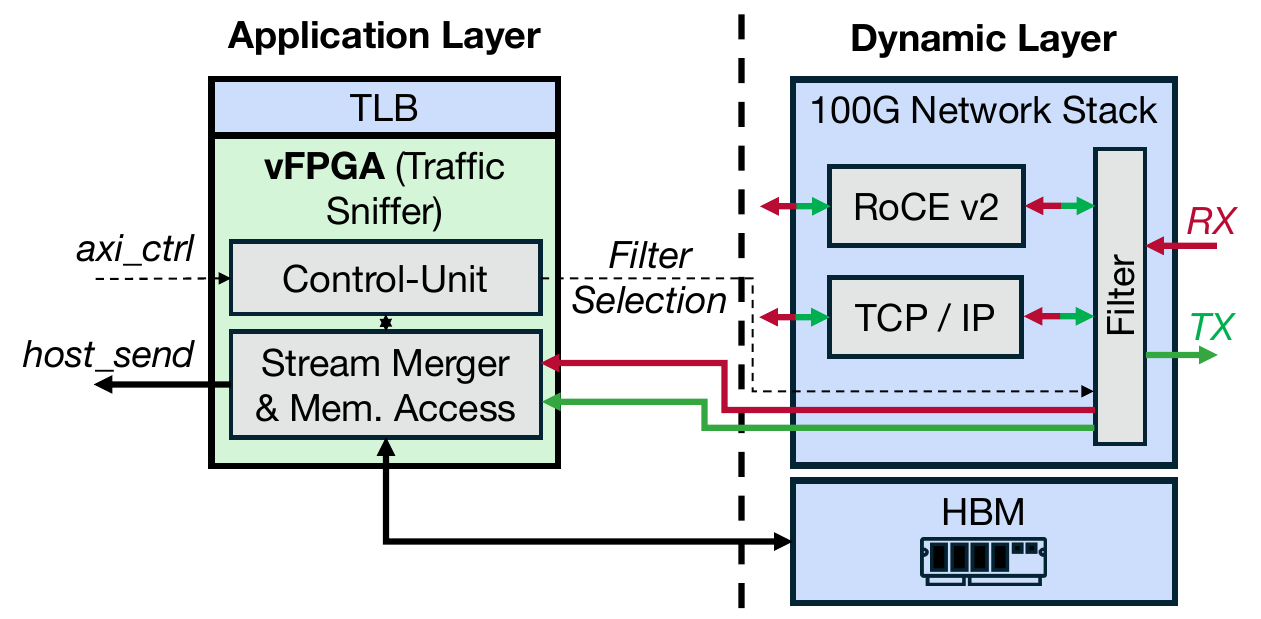}

    \caption{Schematic of the traffic sniffer design with the filter as service and vFPGA-backed application logic.}
    \label{fig:traffic_sniffer_schematic}
\end{figure}

As an example of the design of a service using \coyote, we describe a traffic sniffer (Figure \ref{fig:traffic_sniffer_schematic}) that illustrates the process but also the potential of \coyote for building, e.g., smart NICs. When enabled, a network filter is inserted between the available network stacks (RDMA, TCP/IP) and the 100G CMAC. By utilizing \coyote's control interface and exposing its own registers, the traffic sniffer can be configured from the host software. Hence, RX- and TX-traffic is filtered based on a user-configured filter. Additionally, partial sniffing of only headers is possible through the same control interface. 

On the data plane, the traffic sniffer connects to the shell's networking stacks, the CMAC, and the application layer, which is used to timestamp the data and store it to a previously allocated HBM buffer. With the same control interface, it is possible to start and stop the traffic recording. Using the previously described mechanisms for memory management, the capture data can be synced back to host memory, where a software parser converts the raw packet recordings to a default PCAP file for analysis with standard networking tools, such as \textit{Wireshark}.

In summary, this sniffer design combines key concepts of \coyote's data- and control-flow in the application layer with the notion of a controllable traffic filter as a reconfigurable service to provide a crucial network debugging utility. 

%% file: sections/evaluation.tex
\section{Evaluation}
In the following sections, we evaluate \coyote on a wide range of micro- and macro- benchmarks, showcasing the performance and generalizability of the proposed approach in a realistic data center setting. 

\subsection{Micro-benchmark: Throughput Scaling per App with the Number of HBM Channels}
\label{sec:hbm_results}
\begin{figure}[bt]
    \centering
    \includegraphics[width=\linewidth]{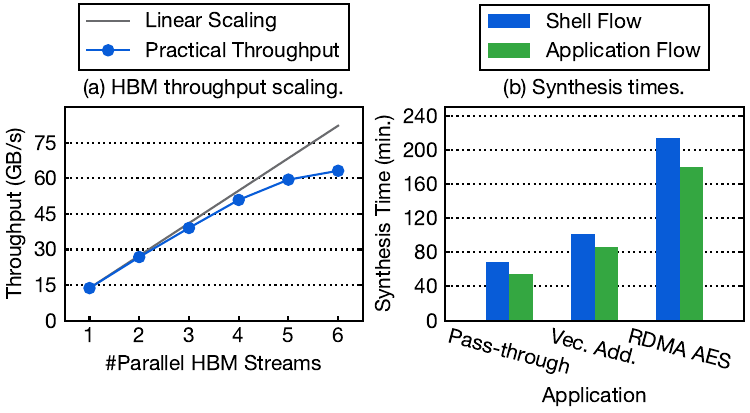}
    \caption{a) Data transfer throughput scaling with number of HBM channels in one vFPGA. b) Comparison of synthesis and implementation times with shell and app flow targetting an Alveo U250.}
    \label{fig:hbm_scaling_synth_times}
\end{figure}

A direct consequence of the improved user interfaces and memory striping is improved memory bandwidth compared to existing shells~\cite{coyote, harmonia}, allowing parallel data transfer and processing in a single vFPGA. Figure~\ref{fig:hbm_scaling_synth_times}(a) illustrates the throughput of a simple pass-through application that consumes data from and stores it back to the card's HBM. The results are  averaged over 50 trials, with 50 warm-up runs and obtained on an Alveo U55C card, with a system clock of 250 MHz and an HBM clock of 450 MHz. Initially, the throughput follows a linear trend, which tapers off as the number of HBM channels increases, due to the memory virtualization overhead. For applications that require the full HBM bandwidth, it is possible to bypass the MMU and directly expose certain HBM channels. While achieving higher performance, this approach would require careful consideration of memory management between the host and the FPGA, which is abstracted by \coyote's virtual memory model. Keep in mind that while the nominal bandwidth on the HBM of an FPGA is higher, it is very difficult to reach it due to design congestion and timing closure issues~\cite{runbin_hbm}.

\subsection{Micro-benchmark: Synthesis Time with Nested Build Flow}
An additional consequence of the three-layer hierarchical hardware design is a decoupled build flow of services and user applications, which speeds up the quite lengthy FPGA synthesis process. For a given shell configuration, it is possible to only synthesize a user application and link it to the shell, without having to resynthesize the dynamic or static layer. As these layers interact with peripherals (PCIe, HBM, CMAC), their synthesis often takes long due to congestion and routing complexity. For example, a cloud provider may want to explore various encryption cores for their RDMA stack; however, having to wait 4-6 hours each time for the synthesis, place and route of the RDMA stack with the encryption core would significantly harm productivity. Instead, the cloud provider could compile the RDMA stack once and link the various encryption cores to it. Figure~\ref{fig:hbm_scaling_synth_times}(b) highlights the synthesis times of the following configurations:
\begin{itemize}
    \item A simple data pass-through application, moving data from one host buffer to another. The only included service is a single data interface to the host.
    \item A vector addition kernel that pulls data from the card's memory and stores the result back to memory. This shell exhibits some synthesis complexity, due to the complexity of memory controllers~\cite{runbin_hbm}.
    \item A shell with RDMA and an AES encryption module. Increased synthesis complexity, due to the presence of a networking stack, which also relies on card memory to handle re-transmissions. 
\end{itemize}

\begin{table}[t]
\caption{Reconfiguration latency for various \coyote shell configs. Average latency with STD reported from 5 trials.}
\begin{tabular}{lccc}
Scenario & \begin{tabular}[c]{@{}c@{}} Coyote kernel \\ latency {[}ms{]}\end{tabular} & \begin{tabular}[c]{@{}c@{}} Coyote total \\ latency {[}ms{]}\end{tabular} & \begin{tabular}[c]{@{}c@{}} Vivado \\ flow [ms] \end{tabular} \\ \hline
\#1 & 51.6 $\pm$ 0.0 & 536.2 $\pm$ 5.2 & 55922.5 $\pm$ 443.1 \\
\#2 & 72.3 $\pm$ 0.2 & 709.0 $\pm$ 9.1 & 63045.2 $\pm$ 78.0 \\
\#3 & 85.5 $\pm$ 0.7 &  929.1 $\pm$ 3.5 & 71417.9 $\pm$ 512.0\\
\end{tabular}
\label{tab:reconfig_times}
\end{table}

Shell flow refers to a flow which synthesizes, places and routes both the application and the services. App flow refers to a flow which only synthesizes, place and routes the user application, which is then linked against a previously routed and locked shell with the target services and placeholders for the application. The time is reported from the start of the build process, including synthesis, placement, routing and any additional checks and optimizations. Overall, the app flow can reduce the synthesis time by 15\% to 20\%.

\subsection{Micro-benchmark: Shell Reconfiguration}
Next, we evaluate \coyote's ability to efficiently reconfigure the entire shell; that is both services and user applications. To do so, we consider three reconfiguration scenarios:
\begin{itemize}
    \item Scenario \#1: Initially, the FPGA is loaded with a simple pass-through kernel and an MMU supporting 2MB pages. Then, the shell is reconfigured to include the same pass-through kernel, but with an MMU supporting 1GB pages.
    \item Scenario \#2: Initially, the FPGA is loaded with a shell supporting RDMA and a single kernel which writes the received traffic to host memory. Then, the shell is reconfigured to have two numerical kernels (vector addition, product) and no networking stack.
    \item Scenario \#3: Initially, the FPGA is loaded with a shell that has both RDMA and the previously mentioned traffic sniffer. Then, it is reconfigured to disable the traffic sniffer, while keeping RDMA enabled.
\end{itemize}

The three scenarios effectively capture \coyote's flexibility and modularity that enable it be applied in a wider range of settings, such as varying the memory model based on the workload (\#1), dynamically increasing the number of user applications (\#2) or enabling/disabling a network debugger (\#3). The results are presented in Table~\ref{tab:reconfig_times}. Since the shell bitsream must be read from disk and copied into kernel space, we report two latencies: the kernel latency, corresponding only to the actual reconfiguration, and the total latency, which includes reading from disk and copying the buffer into kernel space. Additionally, the results include a comparison against a full FPGA re-programming with Vivado Hardware Manager, which also includes a PCIe hot-plug and driver re-insertion.

Partial FPGA reconfiguration still remains a slow process; however, compared to a full re-programming the FPGA, \coyote's shell reconfiguration flow is an order of magnitude faster, even when reading the bitstream from disk. In a realistic setting, shell reconfiguration is expected to happen for significant work-load changes, rather than single application changes, and as such, the latency can be considered acceptable. Since bitstreams are not too large (tens of MBs), the latency penalty incurred for reading from disk, can be solved by keeping certain frequently used shell bitstreams in memory.

\subsection{Macro-benchmark: Multi-tenant AES ECB Encryption}
\begin{figure}[bt]
    \centering
    \includegraphics[width=\linewidth]{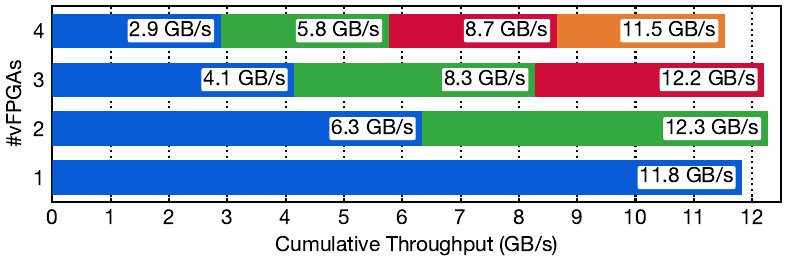}
    \caption{AES ECB bandwidth sharing across vFPGAs.}
    \label{fig:aes_sharing}
\end{figure}

For the next benchmark, we consider deploying multiple instances (vFPGAs) of the AES Electronic Codeblock (ECB) algorithms. While not the strongest encryption algorithm, this benchmark is a valuable test of \coyote's ability to fairly deploy multiple independent applications. Since the algorithm is memory-bound, \coyote must ensure fair sharing of the host memory bandwith (around 12GBps on the Alveo U55C with an XDMA core). As shown in Figure~\ref{fig:aes_sharing}, the bandwidth is indeed fairly distributed accross vFPGAs. Additionally, the cumulative throughput remains constant, indicating no overhead from the hardware arbiter and packetizer.

\subsection{Macro-benchmark: Multi-threaded AES CBC Encryption}
\label{sec:mt_results}

\begin{figure}[bt]
    \centering
    \includegraphics[width=\linewidth]{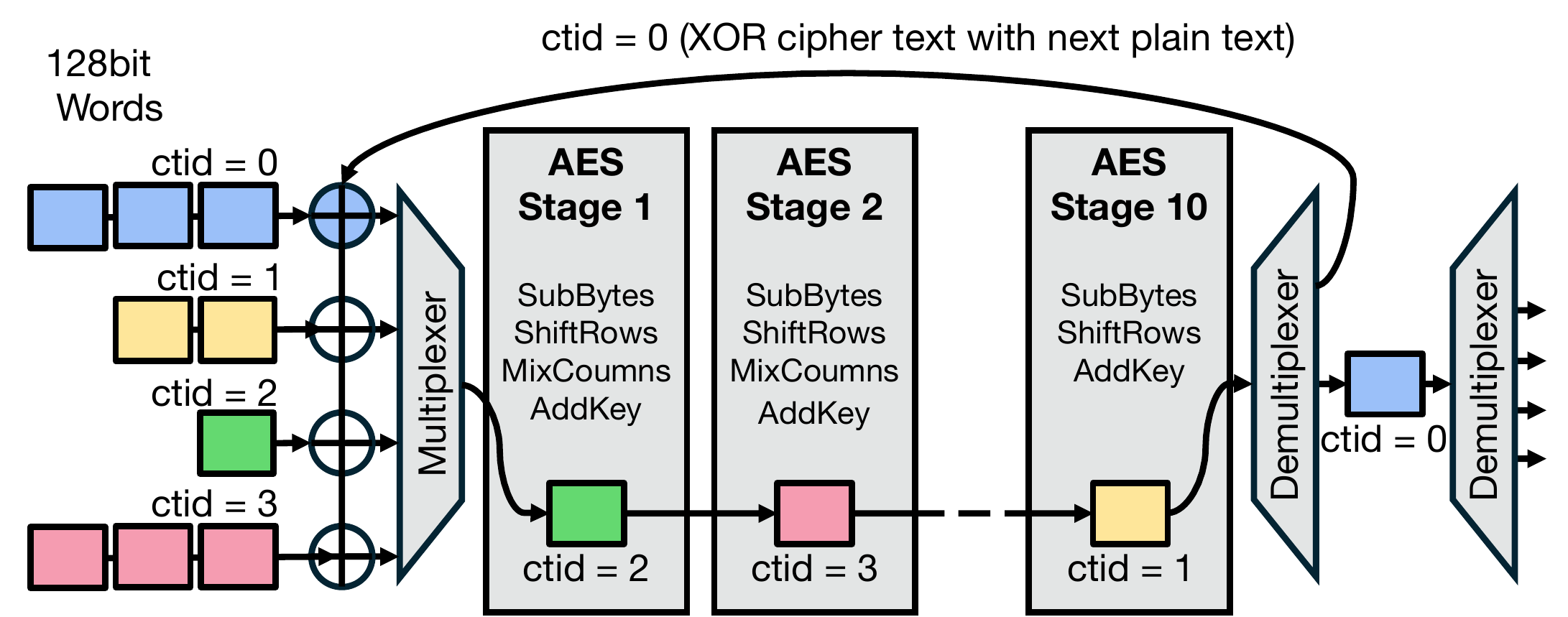}
    \caption{AES CBC multi-threading pipeline.}
    \label{fig:aes_pipeline_example}
\end{figure}

\begin{figure}[b]
    \centering
    \includegraphics[width=\linewidth]{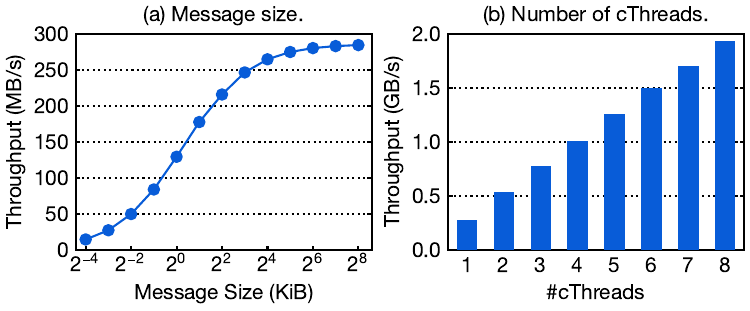}
    \caption{a) Throughput scaling for AES CBC with the message size for a single cThread b) Throughput scaling for AES CBC with multiple cThreads for a 32KB message.}
    \label{fig:aes_results}
\end{figure}

In this benchmark, we consider an alternative encryption algorithm, AES CBC. In CBC mode, the encryption is inherently sequential: each 128-bit text is XOR'ed with the previously encrypted block, leading to pipeline stalls when processing a single thread. Illustrated in Figure~\ref{fig:aes_pipeline_example}, the AES core we use consists of a 10-stage pipeline, which, in a single-threaded execution model, would lead to 9 out of 10 stages remaining idle. Figure~\ref{fig:aes_results}(a) shows that with a single thread, the throughput saturates at 280MBps around 32 KB.

To mitigate this under-utilization, we leverage multiple concurrent software threads (cThreads), associating each request with a unique thread ID. All threads use the same (single) vFPGA on the hardware. Given that data is transferred via AXI streams, this ID is easily assigned to the TID field from the AXI specification. Furthermore, independent processes can submit their text independently, by targeting one of the $N$ host streams. With a single data stream, multi-threading would be more difficult to implement and would require interleaving the chunks of text from different clients in software, also further reducing data isolation. This interleaving would have to be extremely fine-grained, since AES operates on 128-bit chunks, making it memory-inefficient. Additionally, as described in Section~\ref{sec:user_layer}, \coyote transfers data in 512-bit chunks; therefore, packing 128-bit chunks from different clients (IDs) into a single transfer would require additional metadata to be transferred from software to hardware. However, with \coyote's approach, only a simple round-robin arbiter in hardware (pre-provided by \coyote) is needed to select the next encryption input. 

As shown in Figure~\ref{fig:aes_results}b, the throughput scales linearly with the number of software threads, indicating negligible overhead from arbitration and effective reduction of hardware idle time. 

\subsection{Macro-benchmark: HyperLogLog cardinality estimation with on-demand reconfiguration}
\begin{figure}[bt]
    \centering
    \includegraphics[width=\linewidth]{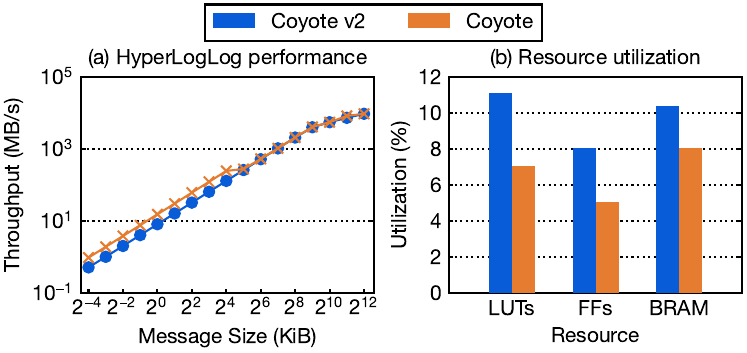}
    \caption{HyperLogLog performance and resource utilization with comparison to Coyote v1.}
    \label{fig:hll_results}
\end{figure}

In this benchmark, we deploy an High-Level Synthesis (HLS) kernel for HyperLogLog (HLL) cardinality estimation~\cite{hll_app} with \coyote, showcasing \coyote's ease-of-integration with high-level programming frameworks as well as applicability to realistic workloads. As a baseline, we use Coyote~\cite{coyote} running the same kernel and compare the achieved throughput and overall (base shell + HLL kernel) resource utilization. Figure~\ref{fig:hll_results} shows that \coyote achieves comparable performance to Coyote, with slightly higher resource utilization. The increased resource utilization is due to the added features to the shell which provide better interfaces and support at no performance cost; however, it is important to note that the overall utilization remains low, around 10\%, leaving ample room for deploying other applications. As an example of the difference in functionality, in \coyote we can run the same kernel as a background daemon loaded on demand. When a client (local or remote) submits a request to run HLL, \coyote loads the kernel through partial reconfiguration and runs it. On average, the partial reconfiguration to load the HLL kernel takes only 57ms. 

\subsection{Macro-benchmark: Neural network inference}
As a final example, we deploy a neural network on the FPGA using \coyote and hls4ml~\cite{Duarte_hls4ml, GitHub_hls4ml}. hls4ml is a widely used, open-source framework that compiles high-level neural networks into quantized FPGA IP cores for real-time inference. Additionally, it includes a set of \emph{accelerator} backends that integrate the generated IP core with the necessary infrastructure for performing inference on PCIe-attached FPGAs. In a similar manner, we extend hls4ml with a new \emph{accelerator} backend, \texttt{CoyoteAccelerator}, which will integrate the generated neural network IP as a vFPGA in \coyote. Due to \coyote's modular design and high-level software abstractions, we were able to seamlessly integrate the backend in hls4ml's Python library, as shown in the following code snippet. Important to note, these changes are completely hidden from end-users of hls4ml: to use the new backend, it is sufficient to simply pass the correct backend name when synthesizing the model. Performing inference is equally simple, and can be achieved through the high-level Python function \texttt{predict}. This example shows how \coyote provides an interaction similar to that found on GPUs, through frameworks such as PyTorch~\cite{pytorch} or TensorFlow~\cite{tensorflow}.

\vfill\eject

\begin{code}
\captionof{listing}{Neural network inference with \coyote and hs4ml.}
\begin{minted}{python}
# Load TensorFlow/Keras model and dataset
model = load_model('sample_keras_model.h5')
X = np.load('sample_data.npy')

# Create hls4ml model targetting Coyote backend
hls_config = config_from_keras_model(keras_model)     
hls_model = convert_from_keras_model(
    keras_model, hls_config=hls_config,
    output_dir='/path/to/output/dir', 
    backend='CoyoteAccelerator', clock_period=4,
    input_data_tb='sample_data.npy',
    output_data_tb=f'sample_labels.npy'
)

# Compile and run software emulation
hls_model.compile()
pred_emu = hls_model.predict(X)

# Start hardware synthesis
hls_model.build()

# Once done, create an "Overlay" of the vFPGA
overlay = CoyoteOverlay('/path/to/output/dir')
overlay.program_fpga()

# Run inference on hardware
pred_fpga = overlay.predict(X, (1,), BATCH_SIZE)
\end{minted}
\end{code}

\begin{figure}[bt]
    \centering
    \includegraphics[width=\linewidth]{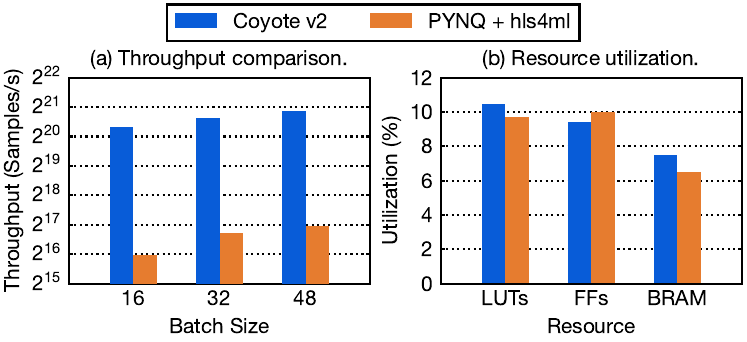}
    \caption{hls4ml performance and resource utilization with \coyote compared to \texttt{PYNQ} with Vitis.}
    \label{fig:hls4ml_results}
\end{figure}

We evaluate our backend against the hls4ml baseline, which uses the Vitis compilation flow with \texttt{PYNQ} for the Python interface. We deploy a neural network for network intrusion detection~\cite{quantized_unsw_nisd, finn_unsw_nisd}, a common use case for FPGAs, on a Alveo U55C clocked at 250MHz. It is important to note that the \texttt{CoyoteBackend} is not tied to a specific model: the actual model conversion and IP generation are the responsibility of the core hls4ml compiler, and any model that is supported by hls4ml can be deployed with \coyote. The results are shown in Figure~\ref{fig:hls4ml_results}, indicating a clear advantage in performance of the proposed backend, while keeping the overall resource utilization approximately equal. While there is a clear advantage in performance, it is important to note that the baseline is not fully optimized, since it requires the data to be copied from host memory to FPGA HBM, before being consumed by the neural network, rather than being streamed directly into the model from the host. Part of the slow-down comes from the fact that the \texttt{CoyoteBackend} integrates directly with \coyote's high-performance C++ library, whereas \texttt{PYNQ} provides a number of additional features and control steps for FPGAs, implemented in Python. 

%% file: sections/conclusion.tex
\section{Conclusion}
In this paper we have described \coyote, an open-source FPGA shell with a novel hierarchical architecture and a unified user interface that enables support for reconfigurable services and hardware multi-threading. The proposed approach reduces synthesis and reconfiguration times, while enabling seamless deployment of arbitrary applications (e.g. HyperLogLog, encryption, neural networks). In future work, we want to extend \coyote as a platform for heterogeneous systems, add support for services such as collective communication~\cite{accl}, interaction with storage systems, and improved high-level APIs for ML applications (e.g., PyTorch~\cite{pytorch}).